\documentclass[twocolumn,prl,superscriptaddress,showpacs,byrevtex]{revtex4}
\usepackage{epsf,graphicx,amssymb}
\usepackage[usenames]{color}
\usepackage{natbib}
\usepackage{float}
\usepackage{gensymb}
\usepackage{amsmath,amssymb}
\usepackage{color}

\begin{document}

\title{Observation of the resonance frequencies of a stable torus of fluid}

\author{Claude Laroche}
\affiliation{Laroche Laboratory, Rue de la Madeleine, F-69 007 Lyon, France}
\author{Jean-Claude Bacri}
\affiliation{Universit\'e de Paris, Universit\'e Paris Diderot, MSC, UMR 7057 CNRS, F-75 013 Paris, France}
\author{Martin Devaud}
\affiliation{Universit\'e de Paris, Universit\'e Paris Diderot, MSC, UMR 7057 CNRS, F-75 013 Paris, France}
\author{Timoth\'ee Jamin}
\affiliation{Universit\'e de Lyon, ENS de Lyon, CNRS, Lab. de Physique  \& UPMA, F-69342 Lyon, France}
\author{Eric Falcon}
\email[E-mail: ]{eric.falcon@univ-paris-diderot.fr}
\affiliation{Universit\'e de Paris, Universit\'e Paris Diderot, MSC, UMR 7057 CNRS, F-75 013 Paris, France}

\date{\today}
\begin{abstract}We report the first quantitative measurements of the resonance frequencies of a torus of fluid confined in a horizontal Hele-Shaw cell. By using the unwetting property of a metal liquid, we are able to generate a stable torus of fluid with an arbitrary aspect ratio. When subjected to vibrations, the torus displays azimuthal patterns at its outer periphery. These lobes oscillate radially, and their number $n$ depends on the forcing frequency. We report the instability ``tongues'' of the patterns up to $n=25$. These resonance frequencies are well explained by adapting to a fluid torus the usual drop model of Lord Rayleigh. This approach could be applied to the modeling of large-scale structures arisen transiently in vortex rings in various domains.


\end{abstract} 


\maketitle


Vortex rings are ubiquitous in Nature. They occur at different scales in various domains: hydrodynamics \cite{Maxworthy72}, plasma physics \cite{Gharib17,CothranPRL09}, geophysics during volcanic eruptions \cite{TaddeucciGRL15}, quantum gravity \cite{Abramowicz06}, and biophysics such as bioconvection \cite{Simkus09}, or underwater bubble rings produced by dolphins \cite{Marten96}. Since their first mathematical analysis in 1858 \cite{Helmholtz58}, their formation, dynamics, or collision have been extensively studied \cite{Batchelor67,Saffman92,Lim92}. Indeed, it is common to generate transient vortex rings (such as smoke rings) in laboratory by pushing a fluid out of a tube or of a hole \cite{Walker87,Saffman78}, by impacting a solid disk in a fluid at rest \cite{Deng17} or a droplet on a plate \cite{RenardyJFM03} or in another liquid \cite{BaumannPoF98,SostareczJFM03}, or by vibrating a gas bubble in a liquid \cite{Zoueshtiagh06}. Once the external force vanishes, the fluid ring is inevitably unstable, and breaks up in droplets \cite{Worthington80,Pairam09,Bird10,McGraw10,FragkopoulosPRE17}. The dynamics of a vortex ring has been shown to be dominated by large-scale structures, such as azimuthal modes observed both experimentally \cite{Walker87,Lim92,Deng17} and numerically \cite{OrlandiJFM93,ChengJFM10}. However, the details of the growth mechanisms of such instability are not well understood \cite{Deng17}, reflecting in large measure the experimental difficulties in generating a stable fluid torus under well-controlled conditions. 

Forming a stable ring of fluid is indeed far from being simple and still remains a challenge. For instance, fluid tori can be obtained by levitating a liquid over its vapor film (Leidenfrost effect) \cite{Perrard12} or by injecting a liquid crystal in a rotating bath containing a yield-stress material as outer fluid \cite{PairamPNAS13}. Here, we report on a much simpler technique to generate a stable torus with an arbitrary aspect ratio. We inject an unwetting fluid (mercury) at the periphery of a solid cylinder, and a stable torus of fluid is forming. This solid boundary prevents the undulations of the torus inner periphery that would occur with no confinement to minimize its surface. We then report the first quantitative measurements of the torus resonance frequencies. When subjected to vibrations, azimuthal modes occur at its outer periphery up to $n=25$. We understand our results by adapting to a fluid torus the usual Rayleigh's model valid for a puddle \cite{Lamb,Rayleigh79}. The geometric properties of the section of the torus can be also inferred from a poloidal instability (driven by the confinement) occurring before each azimuthal pattern. Finally, these patterns look like the ones arisen transiently in vortex rings dynamics \cite{Walker87,Deng17,Simkus09}, and our approach could be thus applied to their modeling.




\begin{figure}[!h]
\begin{center}
\includegraphics[scale=0.5]{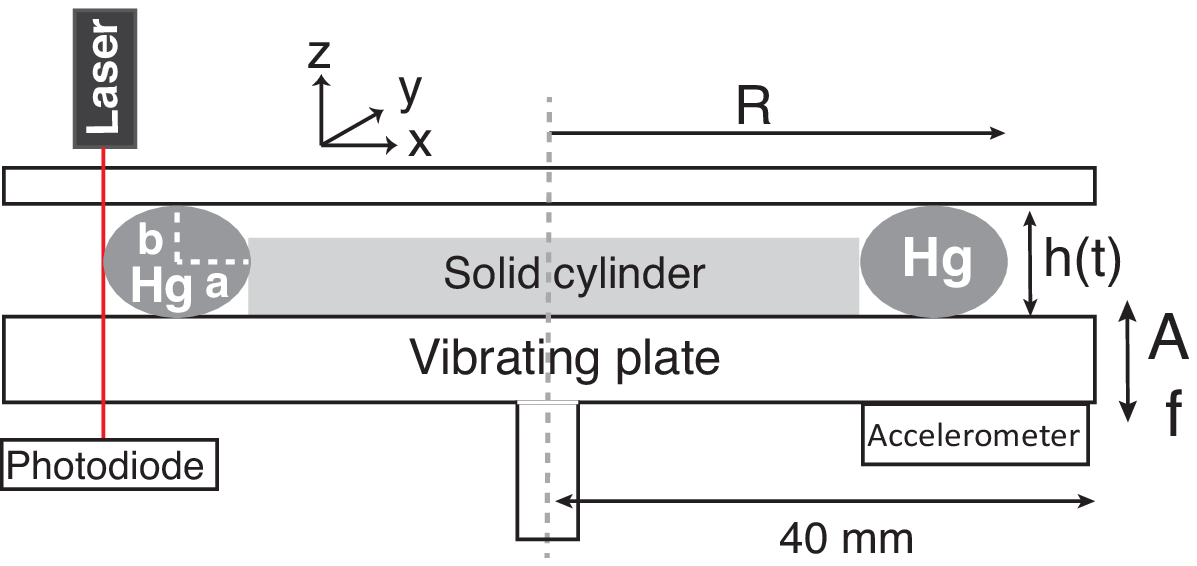}
\includegraphics[scale=0.13]{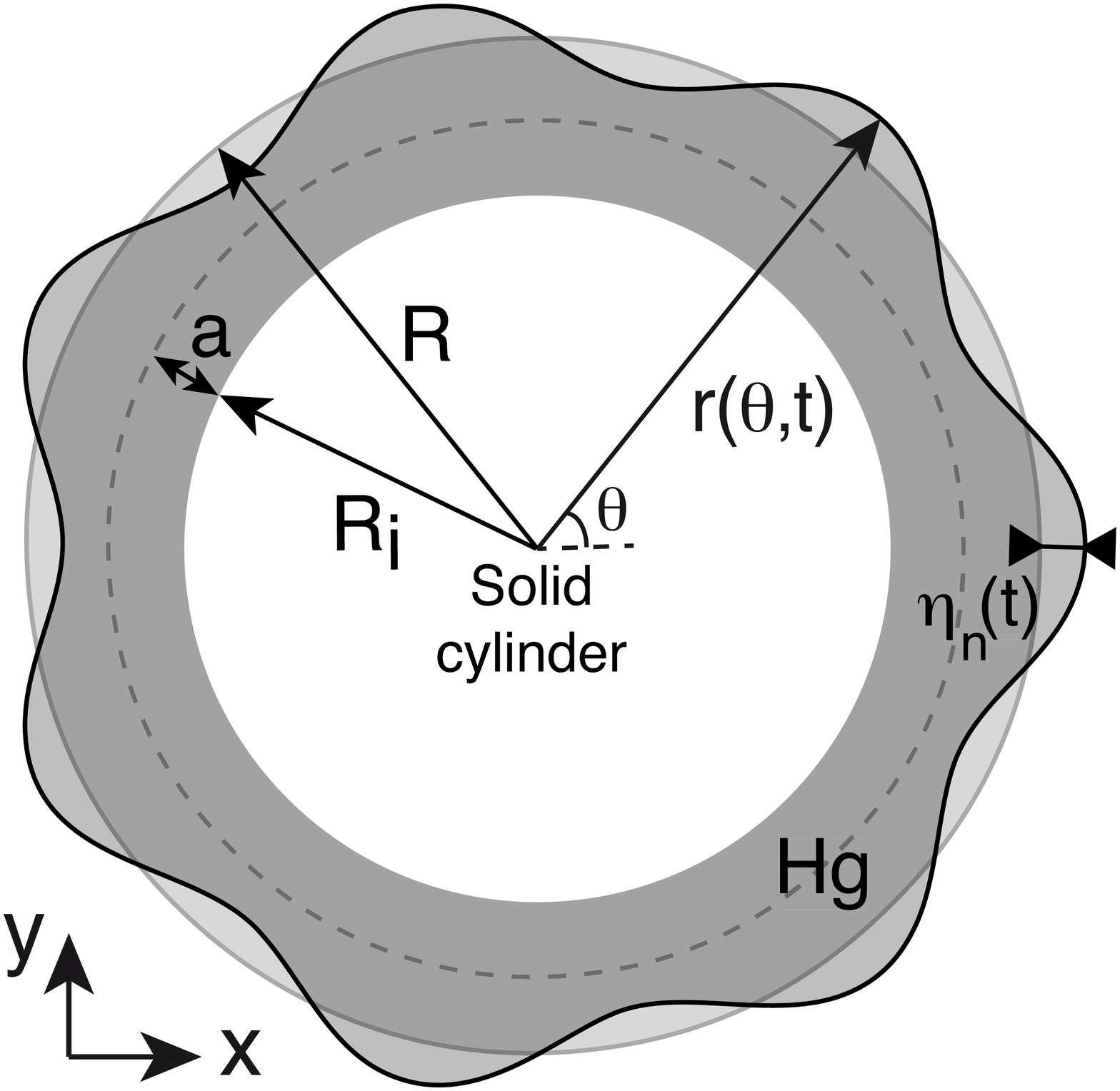}
\caption{Top: Experimental setup. Plate forcing: $A\sin{(2\pi ft)}$. Bottom: top view of fluid torus deformations (mode 7).}
\label{fig01}
\end{center}
\end{figure}


A schematic view of the experimental setup is shown in Fig.\ \ref{fig01}. A volume of mercury ($V=1.5$ mL) is injected, between two transparent plates, at the periphery of a cylinder to form a stable torus of fluid of outer (resp. inner) radius $R=21$ mm (resp. $R_i=15$ mm). The distance between the top and bottom plates is $h=1.5$ mm (except otherwise stated). Due to gravity, the torus is flattened and has an elliptical section of semi-major and semi-minor axes $a=(R-R_i)/2=3$ mm and $b=h/2$ (see Fig.\ \ref{fig01}). The aspect ratio of the torus $\chi\equiv(R+R_i)/(2a)=6$, roughly twice the one of typical doughnut confectionery. The top glass plate (fixed to the laboratory frame of reference - except otherwise stated) avoids possible axisymmetric modes on the top of the ring. The bottom Plexiglass plate of slightly conical shape is coated with a spray to obtain a substrate repelling liquids \cite{UED},
and is subjected to vertical sinusoidal vibrations by means of an electromagnetic shaker (frequency $f \leq 65\,\mathrm{Hz}$ and amplitude $A < 0.5$ mm). An accelerometer (BK 4393) below the plate measures its acceleration ($< 4.9$ m.s$^{-2}$).
The horizontal oscillations of the torus outer periphery are recorded by means of a photodiode measuring the luminous flux, received from a laser, the flux being modulated by the fluid oscillations cutting the laser beam. Direct visualization is obtained with a camera (IDS UI3160CP) located above the drop. Mercury properties are: density, $\rho=13 500\,\mathrm{kg.m^{-3}}$, surface tension, $\gamma=330\,\mathrm{mN.m^{-1}}$, and kinematic viscosity, $10^{-7}\,\mathrm{m^{2}.s^{-1}}$. Its capillary length is $l_c=\sqrt{\gamma/(\rho g)}=1.6\,\mathrm{mm}$. $\gamma$ is measured 32\% lower than the reference one due to the presence of contaminants and possible slight oxidation, even if special attention was paid. Before each experiment, mercury is renewed by using a fresh fluid volume from the bottom of the reservoir, free of contaminants due to the heaviness of mercury. 

\begin{figure}[!t]
\begin{center}
\includegraphics[scale=0.206]{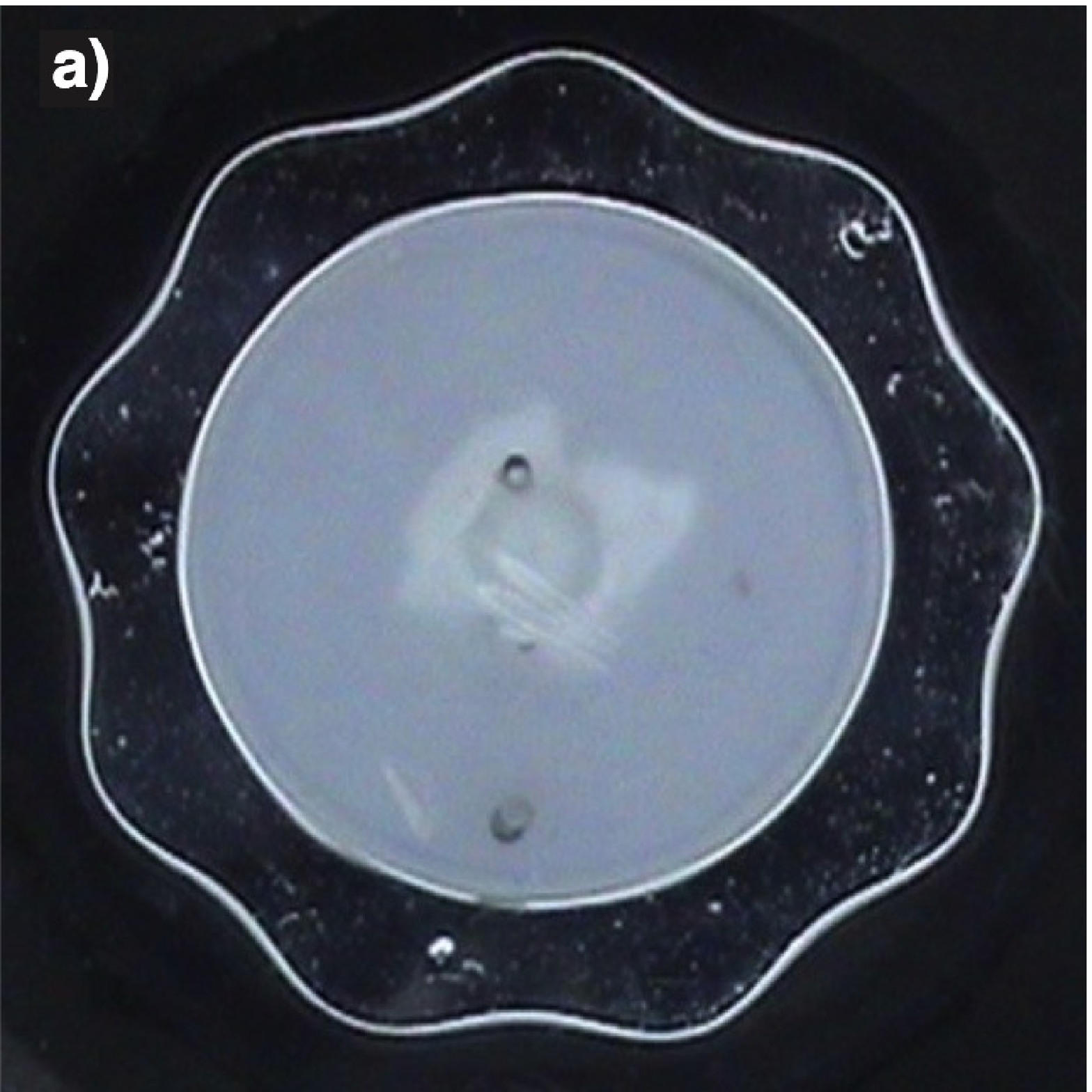}
\includegraphics[scale=0.208]{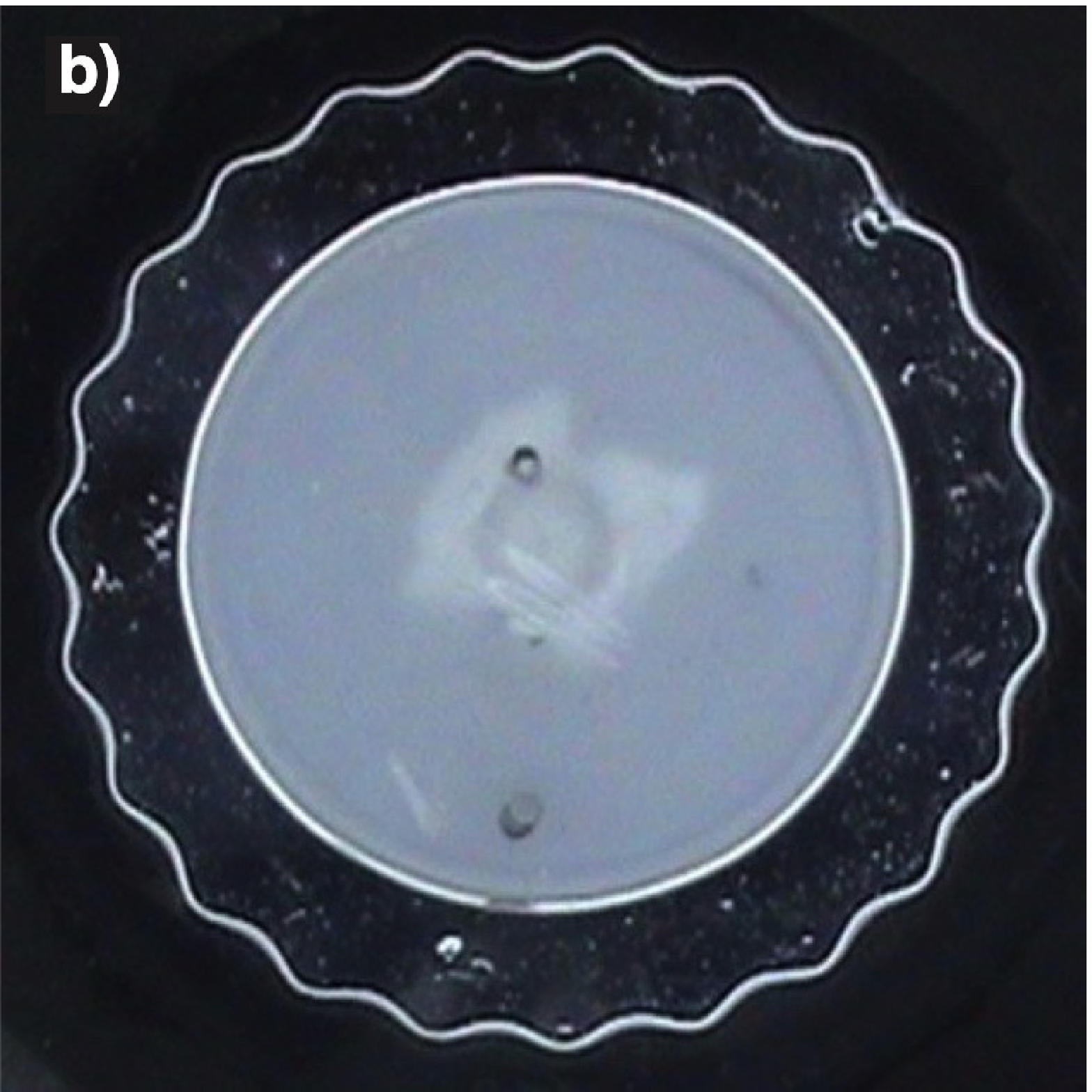}
\includegraphics[scale=0.22]{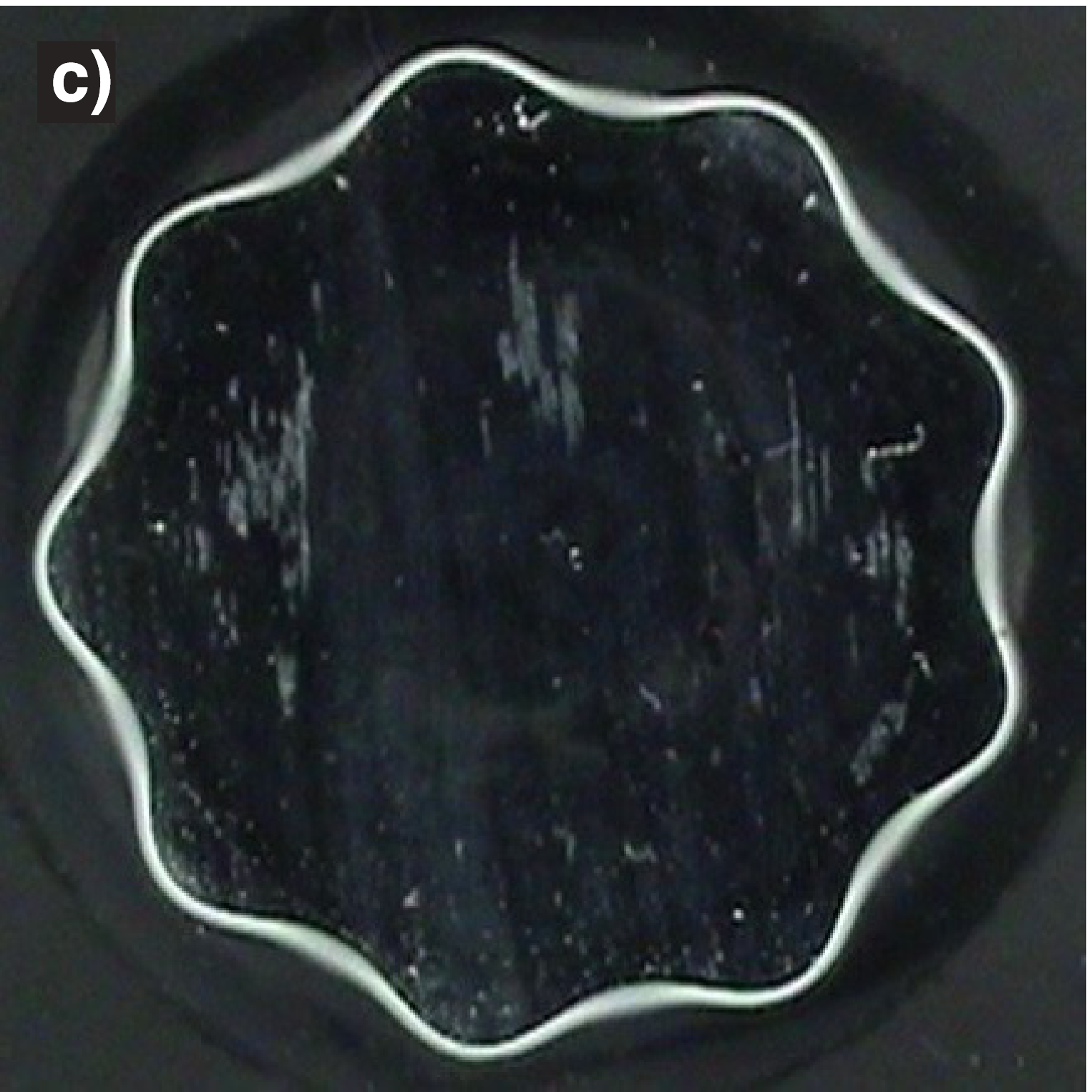}
\includegraphics[scale=0.22]{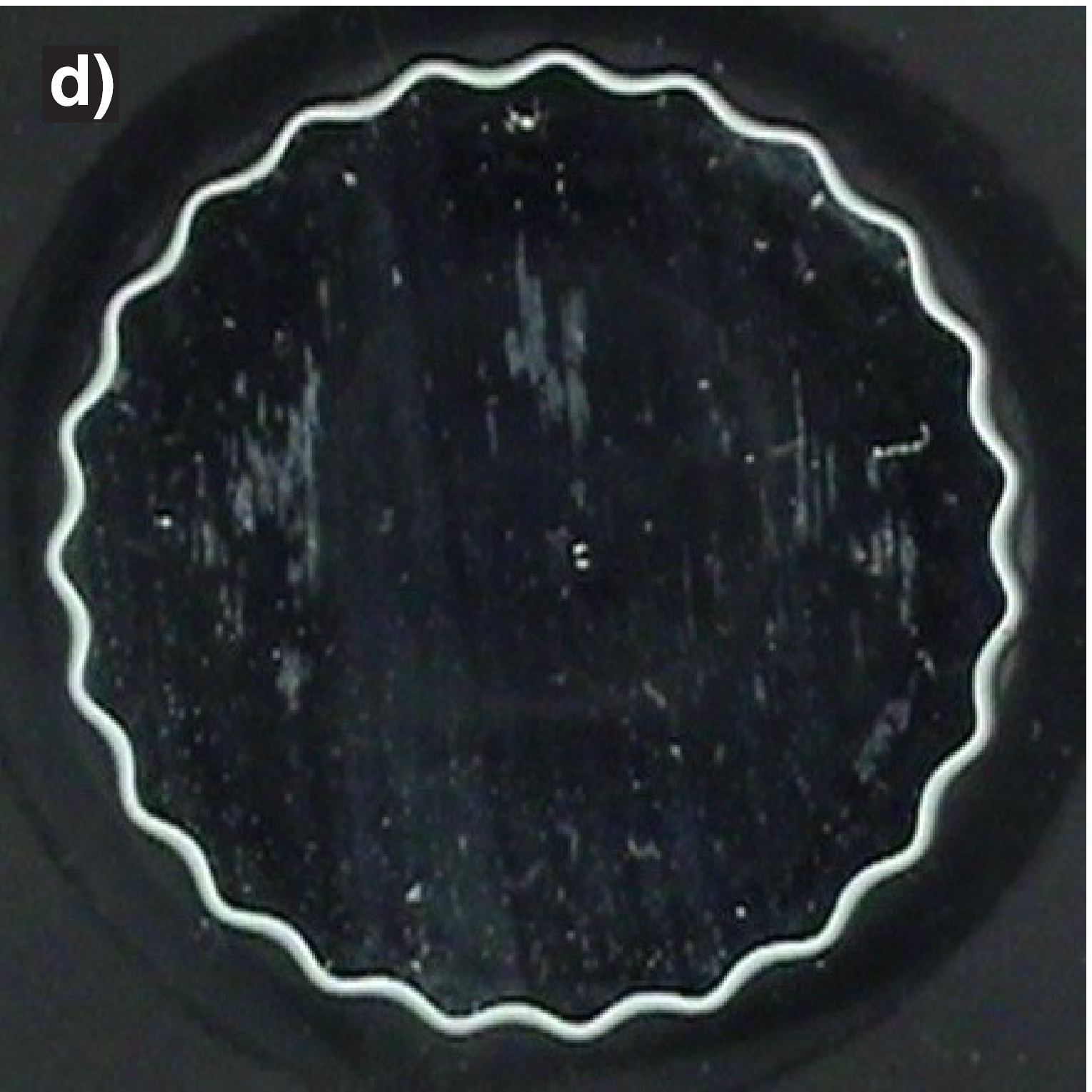}
\caption{Top view of the azimuthal pattern displayed at the periphery of a torus (a, b) and of a puddle (c, d) of mercury for $f=16.4$ Hz (a, c) and $f=43$ Hz (b, d). The corresponding numbers $n$ of lobes, oscillating radially at $f/2$ are $n=9$ (a, c) and $23$ (b, d), respectively. Grey areas correspond to the solid cylinder (a, b). Torus or puddle diameter at rest $\approx 42$ mm. $V=1.5$ mL (a, b). $V=3.2$ mL (c, d).}
\label{fig03}
\end{center}
\end{figure}

To accurately measure the onset of azimuthal oscillations of the torus, we process as follows. The shaker is driven sinusoidally at frequency $f$ with a slowly increasing ramp in amplitude, $A$, the latter being controlled and measured by the voltage across the shaker over time. The photodiode measures the amplitude of the horizontal oscillations of the torus outer periphery over time. By comparing both amplitudes oscillating initially in phase at $f$, one can define accurately the onset of the instability when the fluid oscillations become slightly modulated in amplitude as a precursor of the subharmonic behavior at $f/2$ of the fluid in response to the forcing at $f$ (see Fig. 1 in Supp. Material \cite{SM}). Then, we note the corresponding amplitude of the shaker, $A_c$ and the mode number, $n$, of the pattern observed from the camera. Indeed, above this critical amplitude, $A_c$, an azimuthal pattern is observed in the horizontal plane at the torus outer periphery (see Fig.\ \ref{fig03}a): lobes oscillate radially at $f/2$, i.e. half the forcing frequency. When $f$ is increased, the number $n$ of oscillating lobes increases as shown in Figs.\ \ref{fig03}a-b. Movies and more pictures are shown in Supp. Material \cite{SM}.

Similar experiments are also carried out without the solid cylinder located in the center of the cell, to form a flattened puddle of fluid (see Figs.\ \ref{fig03}c-d) with an outer diameter adjusted to the previous value for the torus. The pattern observed for a torus is found to have the same properties as the one for a puddle, for the same $f$ (see Figs.\ \ref{fig03}a and c or Figs.\ \ref{fig03}b and d). Moreover the evolution of $n$ with $f$ is similar. No azimuthal oscillation in the inner diameter of the torus occurs. This means that the instability occurs only near the ring outer periphery and the bulk plays no major role. 

The derivation of the eigenfrequency $f_n$ of an inviscid fluid torus has to be independent of the nature of the forcing and arises from the interplay between inertia and surface tension effects. We consider small radial deformations of the outer peripheral surface of amplitude $\eta_n(t)\ll R$. In polar coordinates, this reads $r(\theta,t)=R+\eta_n(t)\cos(n\theta)$ (see Fig.\ \ref{fig01}). In the limit $2R \gg h$, the torus shape is approximated by a thin hollow cylinder. The radial amplitude of the lobes $\eta_n(t)$ is then governed by a harmonic oscillator equation, $\mathrm{d}^2\eta_n(t)/\mathrm{d}t^2+\omega_n^2 \eta_n(t)=0$, of eigenfrequency $f_n$ (see Supp. Material \cite{SM})  
\begin{equation}
\omega^2_n= \frac{\gamma}{\rho R^3}n(n^2-1) \frac{1-\left(R_i/R\right)^{2n}}{1+\left(R_i/R\right)^{2n}} \ \ {\rm for}\ \ n>1
\label{fn}
\end{equation}
with $\omega_n=2\pi f_n$. To wit, we have adapted the Rayleigh derivation valid for a puddle ($\omega^2_n= \frac{\gamma}{\rho R^3} n(n^2-1)$ \cite{Lamb,Rayleigh79})  to a fluid torus. A correction factor ($\leq 1$)  thus occurs that tends to 1 for large $n$. For our torus aspect ratio, both models are almost similar (1.3\% difference for $n=2$ and 0.2\% for $n=3$).
In addition to these azimuthal modes, an axisymmetric mode due to the flattening of the confined torus on the top plate can be also derived (see Supp. Material \cite{SM})
\begin{equation}
\omega_1^2=\frac{\gamma}{\rho h R^2} \cdot \frac{2(R^2-R_i^2)}{\left(\frac{R^2-3R_i^2}{4}+\frac{R_i^4}{R^2-R_i^2}\ln\frac{R}{R_i}\right)}
\label{f1}
\end{equation}

The vertical vibrations of the substrate force the fluid parametrically, leading thus to a Mathieu equation for $\eta_n(t)$  for weak enough vibrations \cite{Yoshiyasu96,Jamin16}. The solutions of such a parametric oscillator are marginality curves (or instability ``tongues'') separating stable zones (no drop deformation) and unstable zones where azimuthal standing waves at the drop periphery oscillate at $f/2$, near the resonance frequencies $f_n$ \cite{Mathieu}. This corresponds to the parametric instability shown in Fig.\ \ref{fig03}.  %

\begin{figure}[t]
\begin{center}
\includegraphics[scale=0.4]{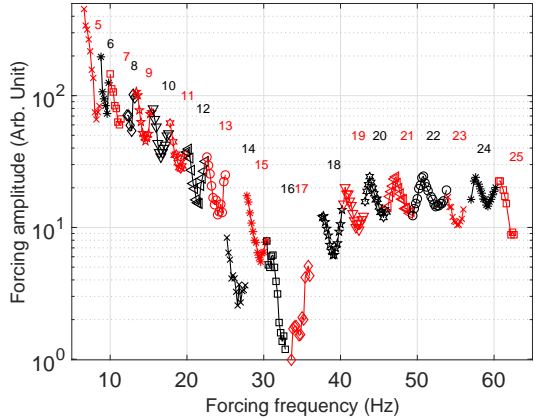} 
\caption{Phase diagram of the lobes occurring on a vibrating torus showing the resonance tongues from $n=5$ to 25. To better distinguish successive tongues, the corresponding data and mode number $n$ are colored alternatively in black and red (light-grey). $V=1.5$ mL.} 
\label{fig04}
\end{center}
\end{figure}

Experimentally, these tongues of instability are displayed in Fig.\ \ref{fig04} as a function of the critical amplitude of vibration, $A_c$, and the forcing frequency $f$. Below each resonance tongue, no significative deformation is observed whereas, above each tongue, $n$ lobes are observed oscillating at the torus outer periphery near $f/2$, as expected theoretically. Indeed, the Mathieu equation predicts that the minimum of each marginality curve occurs at twice the eigenmode $f= 2f_n$. We observe experimentally up to $n=25$ modes whatever the chosen geometry (torus or puddle). Data are obtained for various $f$ by increasing the forcing amplitude until we observe lobes at critical amplitude $A_c$. Note that a slight hysteresis of the tongues is reported if we decrease $f$ after having increased it.

\begin{figure}[t]
\includegraphics[width=8cm]{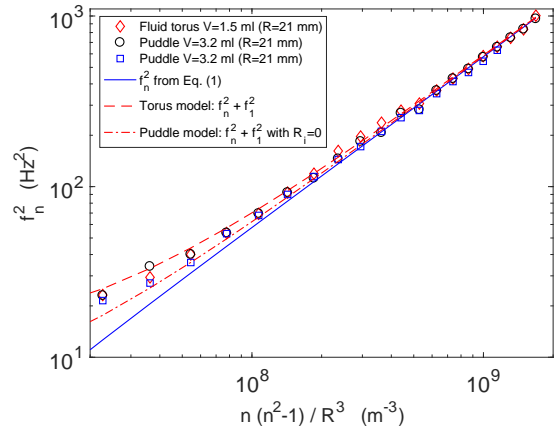} 

    \caption{Resonance frequency $f_n$ of mode $n$ vs $n(n^2-1)/R^3$ for a fluid torus ($\lozenge$) or for two distinct sets of puddle for the same control parameters ($\circ$, $\square$). $R=21$ mm. $n=5$ to~25. Solid line corresponds to Eq.\ (\ref{fn}), dashed and dash-dotted lines to $f^2_n+f^2_1$ [from Eqs.\ (\ref{fn}) and (\ref{f1})] for a torus and a puddle ($R_i=0$),  with $f_1\simeq 3.6$ and 2.1 Hz, respectively.}
\label{fig06}
\end{figure}

\begin{figure*}[!th]
\centering \includegraphics[scale=0.38]{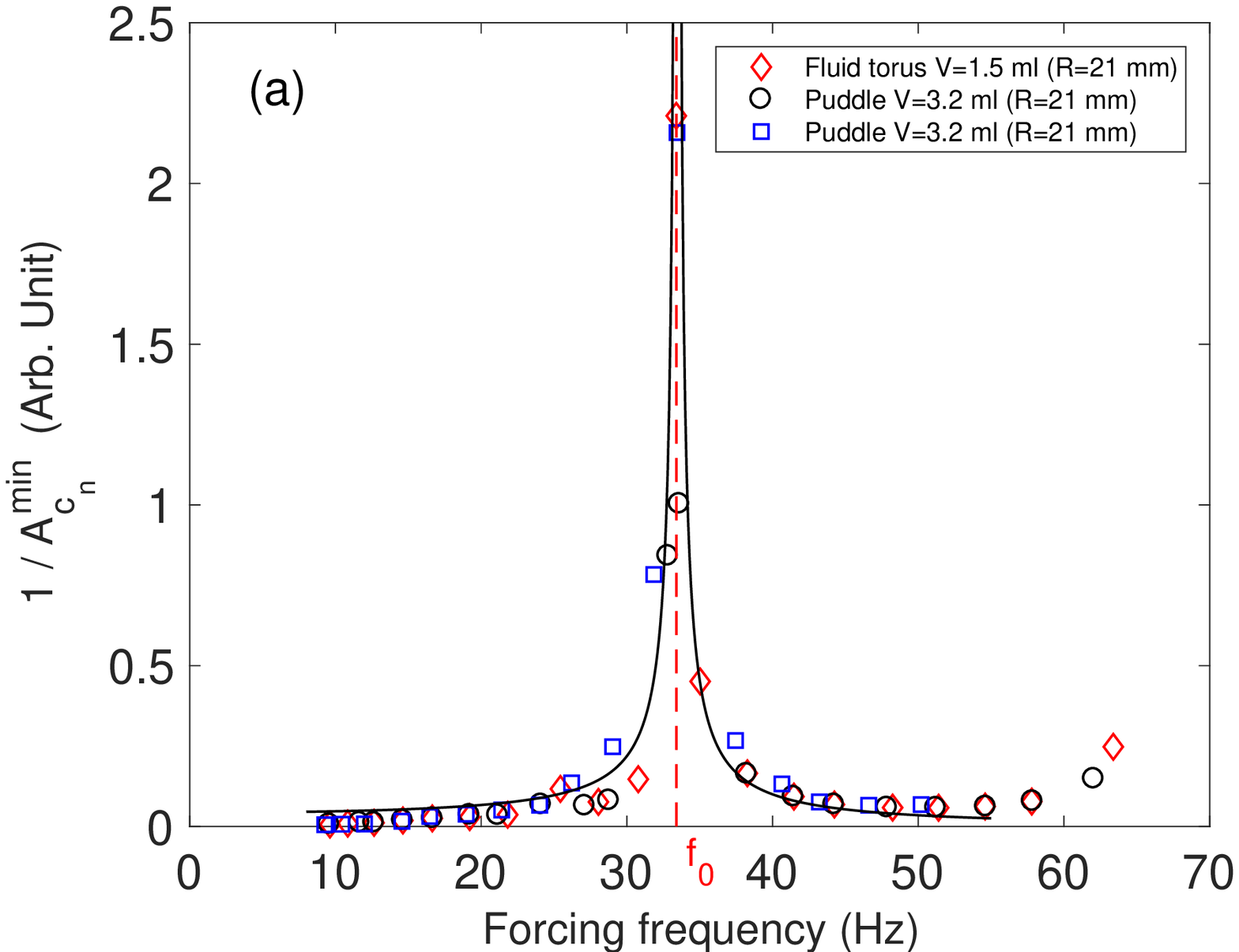}
\includegraphics[scale=0.38]{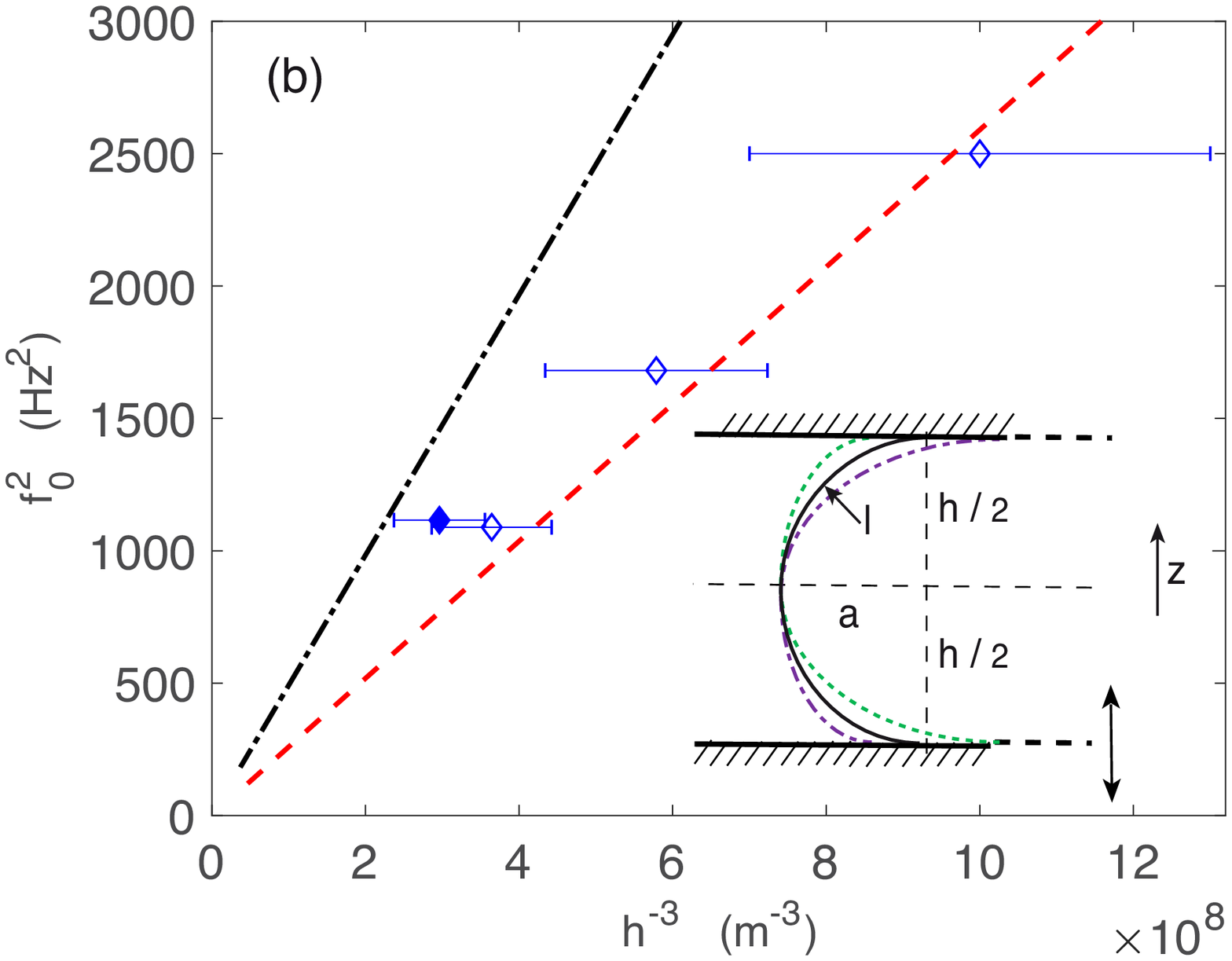}
\caption{(a) Inverse of the minimum amplitude of each tongue vs forcing frequency $f$ for a torus ($\lozenge$) or a puddle ($\circ$, $\square$ - same as in Fig.\ \ref{fig06}). $R\simeq 21$ mm. $h=1.5$ mm. Solid line displays $1/A^{{\rm min}}_{c_n} \sim 1 / \sqrt{(f_0^2-f^2)^2}$ with $f_0^2=2\pi(\gamma/\rho)/(2\alpha h)^3$ and $\alpha=1.71$. (b) Resonance frequency $f_0$ as a function of $h^{-3}$, with $h$ the confinement distance. ($\blacklozenge$) Top plate is fixed whereas bottom one vibrates, same parameters as in (a). ($\lozenge$) Top and bottom plates are fixed together oscillating in phase. $V=2$ mL.  Model with ($--$) $\alpha=1.9$ and ($-\cdot$) $\alpha=\pi/2$. Inset: schematic lateral view of the confined poloidal mode (half-circle case $l=\pi h/2$).}
\label{fig05}
\end{figure*} 

We now define $f^{min,n}$ as the forcing frequency for which the $n$th tongue is minimum in Fig.\ \ref{fig04}.  The resonance frequency $f_n$ of the azimuthal mode $n$ is thus inferred from Fig.\ \ref{fig04} as $f_n=f^{min,n}/2$.  We plot in Fig.\ \ref{fig06} these experimental resonance frequencies for a torus of fluid, and for a puddle, as a function of $n(n^2-1)/R^3$ in order to compare with the prediction of Eq. (\ref{fn}) (see solid line). If one takes into account the flattening axisymmetric mode of Eq. (\ref{f1}) (also called breathing mode for a puddle \cite{MaJFM18}), the agreement with $f^2_n+f^2_1$ is excellent on 2 decades with no fitting parameter and regardless of the fluid geometry. Such a nontrivial coupling would deserve further studies. As shown in the Supp. Material, we reiterate numerous experiments to measure $f_n$  for different experimental parameters (notably $R$, $h$, plates fixed together or not), and the law in $n(n^2-1)/R^3$ is still found to be valid. 

As quantified above, the resonance frequencies of a torus are expected to be the same as the ones of a puddle for large enough $n$. Physically, it occurs when the azimuthal oscillations occurring at the outer boundary of the torus do not feel the presence of the solid central cylinder. This is the case when their typical wavelength is much smaller than the torus horizontal thickness, i.e. $\lambda \ll 4\pi a$. From the torus outer perimeter, one has  $2\pi R\simeq n\lambda$, and thus the condition reads $n \gg R/(2a)=(1+{\rm \chi})/2=3.5$. The mode number $n$ should thus be much larger than the torus aspect ratio. Here, this condition is almost validated since $n \in [5{\rm, \ } 25]$. We cannot reach smaller values of $n$ on Fig.\ \ref{fig04} since the shaker amplitude is limited. Moreover, no break up of the torus by Rayleigh-Plateau (RP) instability is observed here due to the presence of the cylinder. Indeed, RP should occur for $\lambda \geq 2\pi a$, that is for $n\leq R/a=7$.

We now define $A^{{\rm min}}_{c_n}$ as the minimum amplitude of the $n$th tongue in Fig.\ \ref{fig04}. These minima also display a minimum with $f$ (see Fig.\ \ref{fig04}). Figure\ \ref{fig05}a then shows $1/A_c^{{\rm min}}$ as a function of $f$ in the case of a fluid torus or of a puddle. Similar behavior is observed regardless of the geometry: Much less energy is needed to reach the instability threshold near $f=f_0$. This comes from the occurrence of a confined poloidal mode (oscillating at $f$) observed before each azimuthal mode and resonating at $f_0$ [see Supp. Material (Movie a13.mp4) for a puddle]. For different confinement depths $h$, one finds $f_0^2 \sim h^{-3}$, as displayed in Fig.\ \ref{fig05}b, the error bar being due to the depth measurement uncertainty (0.1 mm). This poloidal mode is thus clearly related to the modulation of the confinement and not of the gravity one since $h\lesssim 2l_c$, the acceleration is weak ($< 0.5$g) and the fluid never loses contact with the plates. This poloidal mode can be explained by the oscillations (in phase opposition) of the outer free surface at the top and bottom plates (see inset of Fig.\ \ref{fig05}b). Indeed, assuming the curvilinear length $l$ of the fluid outer boundary to be half a typical wavelength in the vertical plane, $l\sim \lambda_z/2$ (see inset of Fig.\ \ref{fig05}b). One has from geometry $l=\alpha h$, with $\alpha=\pi/2\simeq 1.57$ for a half-circle, or an adjustable parameter experimentally quantifying the more realistic ellipsoidal section of the ring. One has thus $\lambda_z=2\alpha h$. Using the usual dispersion relation of capillary waves $\omega_z^2=(\gamma/\rho)k_z^3$ with $k_z \equiv 2\pi/\lambda_z$, one has thus $f_0^2=2\pi(\gamma/\rho)/(2\alpha h)^3$. Figure \ref{fig05}b shows that this law is in better agreement with experiments for $\alpha=1.9$ (dashed line) than for $\pi/2$ (dash-dotted line). Indeed, the torus is not of circular section due to the fluid-plate interactions. The theoretical frequency response of a simple harmonic oscillator $\sim 1 / \sqrt{(f_0^2-f^2)^2}$, with $f_0$ given above, is then in good agreement with the data, as shown in Fig.\ \ref{fig05}a. 

Finally, an indirect measurement of the geometric properties of the section of the elliptic torus is inferred. Balancing the half-ellipse perimeter $\approx \pi/2\sqrt{2(a^2+b^2)}$ with $l$, and using its eccentricity $e\equiv \sqrt{1-(b/a)^2}$, with $a$ its semi-major and $b=h/2$ its semi-minor axes, one finds $a=h/2\sqrt{8\alpha^2/\pi^2-1}$, and $e=\sqrt{1-[8\alpha^2/\pi^2-1]^{-1}}$. For a circle ($\alpha=\pi/2$), one has $a/b=1$ and $e=0$, as expected. For $\alpha=1.9$, as inferred experimentally from Fig.\ \ref{fig05}b, it corresponds thus to an ellipsoidal section of the ring with $a/b \simeq 1.4$ and $e \simeq 0.7$. These experiments were made with top and bottom plates fixed together, distant of $h$, and vibrating thus in phase (see empty symbols in Fig.\ \ref{fig05}b), in order to avoid the fluid flattening with a fixed top plate and a vibrating bottom one ($\alpha=1.71$). 


In conclusion, we reported the first quantitative measurements of the azimuthal patterns on a stable torus of fluid. Using analytical calculations, we showed that they correspond to the eigenmodes of a thin hollow cylinder provided $n$ is much larger than the torus aspect ratio. This approach could be applied to the modeling of the transient large-scale structures in vortex rings in various domains. Our experimental configuration can be easily used to study a stable vortex ring including (poloidal) vorticity by just applying a Lorentz force to the liquid metal. The solid internal confinement could be also replaced by a toroidal potential. This should reveal more precisely the origin of these transient large-scale structures in vortex rings.

\begin{acknowledgments}
This work was supported by the French National Research Agency (ANR DYSTURB project No. ANR-17-CE30-0004). T. J. gratefully acknowledges the fellowship provided by the programme IMPULSION 2018 from IDEXLYON obtained by Laure Saint-Raymond at ENS de Lyon.
\end{acknowledgments}
\bibliography{goutte2}

\begin{thebibliography}{39}
\bibitem[{Not()}]{Note}
\bibinfo{note}{For the bottom conical plate, the depth between the two plates
  is 2.4 mm at the center, and 1.5 mm at the tore/puddle periphery ($R=21$
  mm).}
\end{thebibliography}
\end{document}